\documentstyle[12pt,lathuile]{article}
\newcommand\as{\alpha_{\mathrm{S}}}
\newcommand\smfrac[2]{{\textstyle\frac{#1}{#2}}}
\renewcommand\d{{\mathrm{d}}}
\newcommand\cO{{\mathcal{O}}}
\begin{document}
\renewcommand{\thefootnote}{\fnsymbol{footnote}}
\begin{titlepage}
\begin{flushright}
  RAL--TR--1999--048 \\ CERN--TH/99--204 \\ hep-ph/9907357
\end{flushright}
\par \vspace{10mm}
\begin{center}
{\mbox{\Large \bf \boldmath
\hspace*{-1cm}The Forward--Backward Asymmetry in NNLO
QCD}~\footnote{Talk given by M. Seymour at 13th Rencontres de Physique
de la Vall\'ee d'Aoste, Results and Perspectives in Particle Physics, La
Thuile, Aosta Valley, Italy, February 28th--March 6th, 1999.
This work was supported in part 
by the EU Fourth Framework Programme ``Training and Mobility of Researchers'', 
Network ``Quantum Chromodynamics and the Deep Structure of
Elementary Particles'', contract FMRX--CT98--0194 (DG 12 -- MIHT).}
\hspace*{-1cm}}
\end{center}
\par \vspace{2mm}
\begin{center}
{\bf Stefano Catani}~\footnote{On leave of absence from INFN,
Sezione di Firenze, Florence, Italy.}\\
\vspace{5mm}
{Theory Division, CERN}\\
{CH-1211 Geneva 23, Switzerland}
\vspace{1cm}\\
{\bf Michael H. Seymour}\\
\vspace{5mm}
{Rutherford Appleton Laboratory, Chilton}\\
{Didcot, Oxfordshire,  OX11 0QX,  England}
\end{center}
\par \vspace{2mm}
\begin{center} {\large \bf Abstract} \end{center}
\begin{quote}
  \pretolerance=1000
  We have recently calculated the second-order QCD corrections to the
  forward--backward asymmetry in $e^+e^-$ annihilation.  Here we recall
  the results and compare them to others in the literature.
\end{quote}
\vspace*{\fill}
\begin{flushleft}
  RAL--TR--1999--048 \\ CERN--TH/99--204 \\ July 1999
\end{flushleft}
\end{titlepage}
\mbox{}\newpage
\renewcommand{\thefootnote}{\arabic{footnote}}
\setcounter{footnote}{0}
\title{THE FORWARD--BACKWARD ASYMMETRY IN NNLO QCD}
\author{Stefano Catani\\
{\em Theory Division, CERN, CH-1211 Geneva 23, Switzerland} \\
{\em (On leave of absence from INFN, Sezione di Firenze, Florence, Italy)}\\
Michael H. Seymour\\
{\em Rutherford Appleton Laboratory, Chilton, Didcot, Oxfordshire,
     OX11 0QX, England}}
\maketitle
\baselineskip=14.5pt
\begin{abstract}
  We have recently calculated the second-order QCD corrections to the
  forward--backward asymmetry in $e^+e^-$ annihilation.  Here we recall
  the results and compare them to others in the literature.
\end{abstract}
\baselineskip=17pt
\newpage

Experimental measurements of the forward--backward and left--right
for\-ward--backward asymmetries in $e^+e^-$ annihilation to fermions
provide some of the best determinations of the weak mixing angle
$\sin^2\theta_{ef\!f}$\cite{electroweak}.  In particular, the
forward--backward asymmetry of $b$ quarks is measured with a precision
of about 2\%, allowing an extraction of $\sin^2\theta_{ef\!f}$ with
almost per mille accuracy.  However, since we are dealing with quarks in
the final state, we must ensure that QCD corrections, both perturbative
and non-perturbative, are understood to at least the same precision.
For the perturbative corrections, this requires working to at least
next-to-next-to-leading order (NNLO).

To date there have been two $\cO(\as^2)$ calculations, both in the
massless approximation and using a slightly different definition of the
asymmetry than the experimental measurements, which use the thrust axis
rather than the quark direction.  The classic calculation of Altarelli
and Lampe\cite{AL} determined the $\cO(\as^2)$ coefficient numerically
and found it to be small.  This result has been the basis of all the
experimental analyses since.  However, the recent analytical calculation
by Ravindran and van Neerven\cite{RvN} obtained a coefficient about
four times bigger.  This discrepancy is comparable to the size of the
experimental errors and needs to be resolved before the final
electroweak fits to the LEP1 data can be made.  The
$\cO(\as^2)$-calculation using the experimentally-used thrust axis
definition, would also be highly desirable.

We have recently performed a numerical calculation of the $\cO(\as^2)$
corrections to the forward--backward asymmetry\cite{CSAfb}.
Anticipating the result, given below, we can say that to the precision
required by experiment we confirm the result of Ravindran and van
Neerven and therefore rule out the result of Altarelli and Lampe.
However, we do have a theoretically-important difference compared to
Ravindran and van Neerven, in that we find that the forward--backward
asymmetry contains terms enhanced by logarithms of the quark mass.  Even
though these terms are numerically tiny for realistic quark masses, as a
point of principle it means that the forward--backward asymmetry of
massless quarks is not perturbatively calculable and non-perturbative
fragmentation functions have to be introduced.

We also calculated for the first time the corrections using the thrust
axis definition rather than the quark direction.  These lie
approximately midway between the results of Refs.\cite{AL} and\cite{RvN}
for the quark axis definition.

We here only briefly sketch the method and give the final result, and
refer the reader to Ref.\cite{CSAfb} for more details.

The simplest definition of the $b$-quark\footnote{Throughout this paper
we explicitly consider the case of the $b$-quark.  The results for the
charm quark can be simply obtained by properly replacing the mass and
couplings of the massive quark.}  forward--backward
asymmetry $A_{FB}$ is
\begin{equation}
\label{afbdef} 
A_{FB} = \frac{N_F - N_B}{N_F + N_B} \;\;,
\end{equation}
where $N_F$ and $N_B$ are the number of $b$ quarks observed in the
forward and backward hemispheres, respectively.

The axis that identifies the forward direction can be defined in a
variety of ways.  In this paper we explicitly consider two different
definitions: the $b$-quark direction, and the thrust axis direction,
which we denote by $A_{FB}^{b}$ and $A_{FB}^{T}$ respectively.

According to the definition in Eq.~(\ref{afbdef}), $A_{FB}$ can be
expressed in an equivalent way in terms of the cross section
\begin{equation}
\label{dcs}
\frac{\d\sigma(e^+e^- \to b + X)}{\d x \;\d\!\cos\theta}
\end{equation}
for inclusive $b$-quark production, where $x$ is the fraction of the
electron energy carried by the $b$ quark and $\theta$ is the angle
between the electron momentum and the direction defining the forward
hemisphere (both energies and angles are defined in the centre-of-mass
frame).

Starting from the distribution in Eq.~(\ref{dcs}), we can introduce the
forward and backward cross sections $\sigma_F$ and $\sigma_B$:
\begin{equation}
  \sigma_{F} \equiv \int_{0}^{1} \d\!\cos\theta \,\int_{0}^{1} \d x \,
  \frac{\d\sigma}{\d x \;\d\!\cos\theta} \;, \;\;\;\; \sigma_{B} \equiv
  \int_{-1}^{0} \d\!\cos\theta \,\int_{0}^{1} \d x \, \frac{\d\sigma}{\d
  x \;\d\!\cos\theta} \;,
\end{equation}
and the symmetric and antisymmetric cross sections $\sigma_S$ and
$\sigma_A$:
\begin{equation}
\label{saxs}
    \sigma_S = \sigma_F + \sigma_B \;, \;\;\;\; \sigma_A = \sigma_F -
    \sigma_B \;.
\end{equation}
We can then write the forward--backward asymmetry as
\begin{equation}
\label{xsratio}
 A_{FB} = \frac{\sigma_A}{\sigma_S} \;\;.
\end{equation} 

In order to calculate this ratio perturbatively, we first separate the
contributions to the cross sections into three classes: flavour
non-singlet $(NS)$, flavour singlet $(S)$, and interference (or
triangle) $(Tr)$ (see Ref.\cite{CSAfb} for their precise definition).
We thus write the cross sections as
\begin{eqnarray}
\label{sdecomp}
\sigma_S &=& \sigma_{S, NS} + \sigma_{S, S}^{(2)} + \sigma_{S, Tr}^{(2)}
+ \cO(\as^3) \;, \\
\label{adecomp}
\sigma_A &=& \sigma_{A, NS} + \sigma_{A, Tr}^{(2)} + \cO(\as^3) \;.
\end{eqnarray}
In this notation, up to $\cO(\as)$ there are only non-singlet
contributions.  Thus, $\sigma_{S, S}^{(2)}, \sigma_{S, Tr}^{(2)}$ and
$\sigma_{A, Tr}^{(2)}$ are proportional to $\as^2$.  There are no
singlet contributions to the antisymmetric cross section $\sigma_A$.

The forward--backward asymmetry is decomposed in a similar way.
Expanding the ratio $\sigma_A/\sigma_S$ up to $\cO(\as^2)$, we write
\begin{equation}
\label{afbdec}
  A_{FB}^{(2)} = A_{FB, NS}^{(2)} +
                 \frac{\sigma_A^{(0)}}{\sigma_S^{(0)}} \left(
                 \frac{\sigma_{A, Tr}^{(2)}}{\sigma_A^{(0)}} -
                 \frac{\sigma_{S, Tr}^{(2)}}{\sigma_S^{(0)}} -
                 \frac{\sigma_{S, S}^{(2)}}{\sigma_S^{(0)}} \right)
                 \;\;,
\end{equation}  
where $A_{FB, NS}^{(2)}$ denotes the non-singlet component:
\begin{equation}
\label{afbns}
   A_{FB, NS}^{(2)} = \frac{\sigma_{A, NS}}{\sigma_{S, NS}} \;.
\end{equation}

The triangle contributions give non-universal (i.e.~non-factorizable)
corrections to both the symmetric and antisymmetric cross sections.
They are calculated in Ref.\cite{AL} for the $b$-quark axis definition
and found to be very small.  To our knowledge their contribution to the
thrust axis definition has never been calculated, but we expect it to be
similarly small.  We therefore neglected it, i.e.~$\sigma_{S, Tr}^{(2)}$
and $\sigma_{A, Tr}^{(2)}$ in Eq.~(\ref{afbdec}), from our calculation.
 
The singlet contribution to the symmetric cross section, $\sigma_S$, is
logarithmically enhanced in the small-mass limit and proportional to
$\as^2\ln^3Q^2/m_b^2$.  An approximate expression for it, denoted by
$F^{{\rm Branco}}$, was used in Ref.\cite{AL}.  It is calculated
exactly to $\cO(\as^2)$ in Refs.\cite{HQ,HQresum}, and
the leading and next-to-leading logarithms are summed to all orders in
$\as$ in Ref.\cite{HQresum}.

In some sense the singlet component is a `background' to the
forward--backward asymmetry measurement and, in fact, in the
experimental analyses (see e.g.~Ref.\cite{abbaneo}) it is
statistically subtracted using Monte Carlo event generators.  We
therefore neglected it, i.e.\ $\sigma_{S, S}^{(2)}$ in
Eq.~(\ref{afbdec}), from our calculation.

Before describing the calculation of $A_{FB, NS}^{(2)}$, we take a
slight diversion to discuss the contribution to it from four-$b$ final
states.  Let us first point out a basic fact.  The four-$b$ process
contributes to both the $b$-quark cross sections $\sigma_S$ and
$\sigma_A$ and the $e^+e^-$ total cross section.  However, they appear
with different multiplicity factors in the two cases.  In the case of
the $e^+e^-$ total cross section the multiplicity factor is simply equal
to unity.  In the contribution to the \emph{inclusive\/} $b$-quark cross
sections $\sigma_S$ and $\sigma_A$, these terms count twice since there
are two $b$ quarks in the final state.  This observation is important in
understanding the results for the non-singlet component of the symmetric
cross section $\sigma_S$ discussed shortly.

After summing and squaring the Feynman diagrams for four-$b$ production,
we obtain two types of contribution: $i)$~those that are identical to
the $b\bar{b}q\bar{q}$ final state but with the other quark $q$ replaced
by an untriggered-on $b$ quark, and $ii)$~those that are genuine
interference terms arising from the fact that the two antiquarks are
indistinguishable, called the $E$-term in Ref.\cite{ERT}.  The squared
diagrams of type~$i)$ are lumped together with the corresponding terms
from $b\bar{b}q\bar{q}$ in the singlet ($\sigma_{S, S}^{(2)}$ in
Eq.~(\ref{sdecomp})), non-singlet ($\sigma_{S, NS}$ and $\sigma_{A, NS}$
in Eqs.~(\ref{sdecomp}) and (\ref{adecomp})) or triangle ($\sigma_{S,
Tr}^{(2)}$ and $\sigma_{A, Tr}^{(2)}$ in Eqs.~(\ref{sdecomp}) and
(\ref{adecomp})) contributions.  The squared diagrams of type~$ii)$,
which give a universal (i.e.~factorizable) correction to both the
antisymmetric and symmetric cross sections, can be considered part of
the non-singlet contributions.

It is not entirely clear how four-quark final states are actually
treated in the different experimental analyses, i.e.~the extent to which
they are genuinely measuring the inclusive cross sections.  Often some
vague statement like ``a four-$b$ final state is more likely to be
tagged than a two-$b$ one, but less than twice as likely'' is made.  To
know what to calculate one must understand the corrections that are
applied for this difference in tagging efficiency, which are not usually
explicitly stated in the papers.  In the absence of a unique
experimental procedure and of a definitive statement from the
experiments on what they are measuring, we make this ambiguity explicit
by multiplying the $E$-term by an arbitrary weight factor $W_E$.  An
inclusive definition would correspond to $W_E=2$ (each $b$ quark
contributing once), while an exclusive definition (the cross section for
events containing at least one $b$ quark) would correspond to $W_E=1$.
Since the forward--backward asymmetry is defined to be the asymmetry of
a differential cross section, it is clear that we must use the {\em
same\/} cross section definition in the numerator and denominator,
i.e.~that $W_E$ must be the same in the symmetric and antisymmetric
cross sections.

Having defined the weight factor $W_E$ for the $E$-term, we can define
the following symmetric and antisymmetric cross sections
\begin{eqnarray}
\label{sxswe}
\sigma_{S, NS}(W_E) &=& \sigma_{S, NS}(W_E=0) + W_E \;\sigma_{S}^{(0)}
\int E_S \;, \\
\label{axswe}
\sigma_{A, NS}(W_E) &=& \sigma_{A, NS}(W_E=0) + W_E \;\sigma_{A}^{(0)}
\int E_A \;,
\end{eqnarray}
where $\int E_S$ and $\int E_A$ denote the integral of the symmetric and
antisymmetric $E$-term, respectively.  We recall that the `truly'
inclusive cross sections in Eq.~(\ref{saxs}) correspond to the
definition with $W_E=2$, i.e.~$\sigma_{S, NS}= \sigma_{S, NS}(W_E=2)$
and $\sigma_{A, NS}= \sigma_{A, NS}(W_E=2)$.

The $\cO(\as^2)$-calculation of the cross sections in
Eqs.~(\ref{sxswe},~\ref{axswe}) and of the corresponding
forward--backward asymmetry in the case of a finite $b$-quark mass is
extremely complicated, and we are not able to perform it.  It is thus
convenient to separate the calculation into a piece that is finite
(although still cumbersome) in the massless limit and a simpler piece
that is not.  Then, the finite piece can be more easily computed in the
massless approximation, while the simpler non-finite piece can be
computed in the massive theory.

It is possible to show\cite{CSAfb} that the inclusive definition, with
$W_E=2$, results in an antisymmetric cross section $\sigma_A$ (or,
analogously, $\sigma_{A, NS}$) that is finite in the massless limit, at
least at $\cO(\as^2)$.  However, in the same limit, the
inclusive symmetric cross section is divergent at $\cO(\as^2)$,
even if we only consider its non-singlet component.  The corrections to
(the non-singlet component of) the forward--backward asymmetry itself
must therefore also be divergent in the massless limit.

This final statement remains true for {\em any\/} value of $W_E>0$.  For
example, with \mbox{$W_E=1$}, the non-singlet part of the symmetric
cross section is finite, but the antisymmetric cross section contains
log\-arithmically-enhanced terms.

The divergences in the non-singlet components correspond to
logarithmi\-cally-enhanced terms $\as^2\ln Q^2/m_b^2$ coming from the
$E$-term in the triple-collinear limit, i.e.~when three fermions of the
four-quark final state become simultaneously parallel.  The integral of
the symmetric $E$-term is calculated numerically in Ref.\cite{CSAfb}
and, neglecting corrections of $\cO(m_b/Q)$, the final result is
\begin{equation}
 \label{ES} \int E_S = C_F\left( C_F- \frac{C_A}{2}
 \right)\left(\frac{\as}{2\pi}\right)^2
 \left[2\left(\frac{13}4-\frac{\pi^2}2+2\zeta_3\right)\ln\frac{Q^2}{m_b^2}
 -8.1790\pm0.0013\right],
\end{equation}
where the analytic coefficient in front of $\ln Q^2/m_b^2$ is
proportional to the integral of the non-singlet Altarelli--Parisi
probability $P_{q {\bar q}}^{NS}(z,\as)$ (see, for instance,
Ref.\cite{CFP}):
\begin{equation}
\int_0^1 \d z \, P_{q {\bar q}}^{NS}(z,\as) = \left( \frac{\as}{2 \pi}
\right)^2 C_F \left( C_F - \frac{1}{2} C_A \right) \left( \frac{13}{4} -
\frac{\pi^2}{2} + 2 \zeta_3 \right) \;,
\end{equation}
and the constant term is the result of our numerical calculation.

Having pointed out that the symmetric $E$-term is divergent in the
massless limit, it is very simple to show how the divergence appears in
the inclusive symmetric cross section.  According to the definition of
the non-singlet component of $\sigma_{S}$, the virtual diagrams that
contribute to $\sigma_{S, NS}$ are exactly those that contribute to the
$e^+e^-$ total cross section.  As for the real diagrams, they only
differ by the contributions coming from the $E$-term.  In the total
cross section, the $E$-term enters with a multiplicity factor $W_E=1$,
and its divergence is cancelled by that of the virtual diagrams.  In the
inclusive $b$-quark cross section, the multiplicity factor of the
$E$-term is $W_E=2$ and, thus, the cancellation of the divergence with
the virtual terms is spoiled.

This argument also allows us to directly compute the
$\cO(\as^2)$-correction to Eq.~(\ref{sxswe}).  Exploiting the fact that
the massless QCD correction to $\sigma_{S, NS}(W_E=1)$ is equal to the
correction $R_{e^+e^-}$ to the total cross section, we write
\begin{equation}
\label{sxswe2}
\sigma_{S, NS}(W_E) = \sigma_{S}^{(0)} \left[ R_{e^+e^-} + (W_E -1) \int
E_S + \cO(\as^3) \right] \;\;.
\end{equation}
Then, we obtain an explicit expression for $\sigma_{S, NS}(W_E)$ by
simply introducing in Eq.~(\ref{sxswe2}) our result in Eq.~(\ref{ES})
for $\int E_S$ and the well-known result\cite{Rhad2} for $R_{e^+e^-}$.
In particular, for the inclusive symmetric cross section we obtain
\begin{equation}
\label{sxsinc}
\sigma_{S, NS} = \sigma_{S, NS}(W_E=2) = \sigma_{S}^{(0)} \left[
R_{e^+e^-} + \int E_S + \cO(\as^3) \right] \;.
\end{equation}

Since both $\sigma_{A, NS}(W_E=2)$ and $\sigma_{S, NS}(W_E=1)$ are
finite when $m_b \to 0$, we can use the dependence on $W_E$ to construct
an unphysical observable that is finite in the massless limit:
\begin{equation}
  \label{AFBfinite} A_{FB}^{(2);\mbox{\footnotesize finite}} \equiv
  \frac{\sigma_{A, NS}(W_E=2)}{\sigma_{S, NS}(W_E=1)} \;.
\end{equation}
This observable is the ratio of the antisymmetric part of the {\em
inclusive\/} cross section and the symmetric part of the {\em
exclusive\/} cross section.  Thus $A_{FB}^{(2);\mbox{\footnotesize
finite}}$ is unphysical in the sense that it is not the
forward--backward asymmetry of a single differential cross section.
Nonetheless the definition in Eq.~(\ref{AFBfinite}) helps us to perform
a massless calculation.  The physical result for $W_E=2$ is then given
by
\begin{equation}
  \label{AFBfull} A_{FB, NS}^{(2)} =
  A_{FB}^{(2);\mbox{\footnotesize finite}} - A_{FB}^{(0)} \int E_S \;,
\end{equation}
where $\int E_S$ is the integral of the symmetric $E$-term, given in
Eq.~(\ref{ES}).

Even in the massless limit numerical two-loop calculations are
prohibitively difficult to set up.  Fortunately there is a cancellation
between the genuinely two-loop effects in the ratio on the
right-hand-side of Eq.~(\ref{AFBfinite}), which does allow its numerical
evaluation.  The total contribution can be written as
\begin{equation}
  \label{ratio2}
  \hspace*{-1cm}
  A_{FB}^{(2);\mbox{\footnotesize finite}} = \frac
  {\sigma_A^{(0)} + \sigma_A^{(1);\mbox{\footnotesize one-loop}} +
  \sigma_A^{(1);\mbox{\footnotesize tree}} + \sigma_A^{(2);\mbox{\footnotesize
  two-loop}} + \sigma_A^{(2);\mbox{\footnotesize one-loop}} +
  \sigma_A^{(2);\mbox{\footnotesize tree}}(W_E=2)} {\sigma_S^{(0)} +
  \sigma_S^{(1);\mbox{\footnotesize one-loop}} +
  \sigma_S^{(1);\mbox{\footnotesize tree}} +
  \sigma_S^{(2);\mbox{\footnotesize two-loop}} +
  \sigma_S^{(2);\mbox{\footnotesize one-loop}} +
  \sigma_S^{(2);\mbox{\footnotesize tree}}(W_E=1)} \;.
  \hspace*{-1cm}
\end{equation}
The $\cO(\as)$-contributions come from the one-loop cross sections
$\sigma^{(1);\mbox{\footnotesize one-loop}}$ for the two-parton process
$e^+e^- \to b {\bar b}$ and the tree-level cross sections
$\sigma^{(1);\mbox{\footnotesize tree}}$ for the three-parton process
$e^+e^- \to b {\bar b} g$.  Similarly the non-singlet
$\cO(\as^2)$-contributions from the two-parton, three-parton and
four-parton final states are denoted by $\sigma^{(2);\mbox{\footnotesize
two-loop}},$ $\sigma^{(2);\mbox{\footnotesize one-loop}}$ and
$\sigma^{(2);\mbox{\footnotesize tree}}$ respectively.  Of course, the
dependence on $W_E$ enters only through the four-parton terms
$\sigma_A^{(2);\mbox{\footnotesize tree}}(W_E=2)$ and
$\sigma_S^{(2);\mbox{\footnotesize tree}}(W_E=1)$.

Each of the cross sections is separately divergent, so they have to be
regularized in some way before being combined together.  In any
regularization scheme that preserves the helicity conservation of
massless QCD\footnote{Note that the relations (\ref{loopcor}) are
explicitly violated for massive quarks.} (for example, dimensional
regularization), we have the properties
\begin{equation}
\label{loopcor}
  \frac{\sigma_A^{(1);\mbox{\footnotesize one-loop}}}{\sigma_A^{(0)}}
  =\frac{\sigma_S^{(1);\mbox{\footnotesize one-loop}}}{\sigma_S^{(0)}},
  \qquad\qquad
  \frac{\sigma_A^{(2);\mbox{\footnotesize two-loop}}}{\sigma_A^{(0)}}
  =\frac{\sigma_S^{(2);\mbox{\footnotesize two-loop}}}{\sigma_S^{(0)}},
\end{equation}
so that if we expand the ratio in Eq.~(\ref{ratio2}) up to
$\cO(\as^2)$, the two-loop corrections cancel, and we obtain
\begin{eqnarray}
  \hspace*{-1cm}
  A_{FB}^{(2);\mbox{\footnotesize finite}} &=& \!\!
  \frac{\sigma_A^{(0)}}{\sigma_S^{(0)}} \Biggl[1+ \Biggl(1-
  \frac{\sigma_S^{(1)}}{\sigma_S^{(0)}} \Biggr)\Biggl(
  \frac{\sigma_A^{(1)}}{\sigma_A^{(0)}} -
  \frac{\sigma_S^{(1)}}{\sigma_S^{(0)}} \Biggr)
  \\\label{afb2exp} && \hspace{-1.5em} +
  \frac{\sigma_A^{(2);\mbox{\footnotesize one-loop}}}{\sigma_A^{(0)}} -
  \frac{\sigma_S^{(2);\mbox{\footnotesize one-loop}}}{\sigma_S^{(0)}} +
  \frac{\sigma_A^{(2);\mbox{\footnotesize tree}}(W_E=2)}{\sigma_A^{(0)}}
  - \frac{\sigma_S^{(2);\mbox{\footnotesize
  tree}}(W_E=1)}{\sigma_S^{(0)}} \Biggr] \,,\;\;\;\nonumber
  \hspace*{-1cm}
\end{eqnarray}
where $\sigma_A^{(1)}$ and $\sigma_S^{(1)}$ are the complete
contributions to the antisymmetric and symmetric cross sections at
$\cO(\as)$, $\sigma^{(1)} = \sigma^{(1);\mbox{\footnotesize one-loop}} +
\sigma^{(1);\mbox{\footnotesize tree}}$.  The first line can be
calculated analytically, but the second line is too complicated to be
able to, so must be done numerically.  Since the two-loop terms have
cancelled, this has the structure of a NLO three-jet calculation, as
first noticed by Altarelli and Lampe\cite{AL}.  Thus the calculation
can be performed using known techniques (we use the dipole-formalism
version of the subtraction method\cite{CS}).

We are finally ready to present our numerical results.  We start with
the unphysical, but finite, quantity defined in Eq.~(\ref{AFBfinite}),
and separate out the different colour factors, as in
Refs.\cite{AL,RvN}:
\begin{eqnarray}
  \label{AFBbCNTdef} A_{FB}^{(2);\mbox{\footnotesize finite};b} &=&
  A_{FB}^{(0)}\left[
  1-\frac{\as}{2\pi}\left(1-\frac{\as}{2\pi}\frac32C_F\right)
  \left(\frac32C_F\right) \phantom{\left(\frac{\as}{2\pi}\right)^2}
  \phantom{\left(\frac{\as}{2\pi}\right)^2}\right.  \nonumber\\&&
  \phantom{\left(\frac{\as}{2\pi}\right)^2}
  \phantom{\left(\frac{\as}{2\pi}\right)^2} \left.
  +\left(\frac{\as}{2\pi}\right)^2C_F\left(
  CC_F+NN_C+TT_RN_f\right)\right] \;,
\end{eqnarray}
with $\as \equiv \as(Q^2)$.  Our numerical results are shown in
Table~\ref{table1}, in comparison with the previous calculations.
\begin{table}
 \begin{center}
  \begin{tabular}{|c|c|c|c|}
    \hline
    $b$-quark axis & $C$ & $N$ & $T$ \\
    \hline
    AL \cite{AL} & $4.4\pm0.5$ & $-10.3\pm0.3$ & $5.68\pm0.04$ \\
    RvN \cite{RvN} & $\frac38 = 0.375$ & $-\frac{123}8 = -15.375$ &
      $\frac{11}2 = 5.5$ \\
    \hline
    \hline
    Our Calculation & $0.3765\pm0.0038$ & $-15.3769\pm0.0034$ &
      $5.5002\pm0.0008$ \\
    \hline
  \end{tabular}
 \end{center}
 \caption{\it Results for the coefficients of the $\cO(\as^2)$ correction to
    the finite part of the forward--backward asymmetry with the $b$-quark
    axis definition, Eqs.~(\ref{AFBfull},~\ref{AFBbCNTdef}).}
 \label{table1}
\end{table}
It is clear that we disagree badly with the results of Altarelli and
Lampe\cite{AL}, but are in excellent agreement with Ravindran and van
Neerven\cite{RvN}, who give the coefficients analytically.  However,
we should recall that $A_{FB}^{(2);\mbox{\footnotesize finite}}$, as
given in Eq.~(\ref{AFBbCNTdef}), is not the forward--backward asymmetry
of a definite cross section.  The physical forward--backward asymmetry
must have subtracted from Eq.~(\ref{AFBbCNTdef}) the
logarithmically-enhanced term of Eq.~(\ref{AFBfull}), which
is not present in the result of Ref.\cite{RvN}.  Thus it seems that
they have somehow computed the unphysical
$A_{FB}^{(2);\mbox{\footnotesize finite}}$ rather than the
forward--backward asymmetry.  In fact, their expression for the
correction to the symmetric cross section ($f_T+f_L$ in their Eqs.~(31)
and~(32)) is actually equal to our $\sigma_{S, NS}(W_E=1)$ in
Eq.~\ref{sxswe2}.  So, the fact that their result for $A_{FB}^{(2)}$
agrees with our $A_{FB}^{(2);\mbox{\footnotesize finite}}$ means that we
confirm their result\cite{RivN,RvN} for the inclusive antisymmetric
cross section $\sigma_A^{(2)}=\sigma_A^{(2)}(W_E=2)$ ($f_A$ in Eq.~(33)
of Ref.\cite{RvN}).

Using our numerical program it is straightforward to calculate the
forward--backward asymmetry with any other axis definition (or cuts, for
example on the value of the thrust).  With the thrust axis definition,
we obtain
\begin{eqnarray}
  \label{AFBTCNTdef} A_{FB}^{(2);\mbox{\footnotesize finite};T} &=&
  A_{FB}^{(0)}\left[
  1-\frac{\as}{2\pi}\left(1-\frac{\as}{2\pi}\frac32C_F\right) \left(1.34
  C_F \right) \phantom{\left(\frac{\as}{2\pi}\right)^2}
  \phantom{\left(\frac{\as}{2\pi}\right)^2}\right.  \nonumber\\&&
  \phantom{\left(\frac{\as}{2\pi}\right)^2}
  \phantom{\left(\frac{\as}{2\pi}\right)^2} \left.
  +\left(\frac{\as}{2\pi}\right)^2C_F\left(
  CC_F+NN_C+TT_RN_f\right)\right],
\end{eqnarray}
with $\as \equiv \as(Q^2)$ and the coefficients given in
Table~\ref{table2}.
\begin{table}[b]
 \begin{center}
  \begin{tabular}{|c|c|c|c|}
    \hline
    thrust axis & $C$ & $N$ & $T$ \\
    \hline
    \hline
    Our Calculation & $-3.7212\pm0.0065$ & $-9.6011\pm0.0049$ &
      $4.4144\pm0.0006$ \\
    \hline
  \end{tabular}
 \end{center}
 \caption{\it Results for the coefficients of the $\cO(\as^2)$ correction to
    the finite part of the forward--backward asymmetry with the thrust
    axis definition, Eqs.~(\ref{AFBfull},~\ref{AFBTCNTdef}).}
 \label{table2}
\end{table}
The logarithmically-enhanced piece that has to be added to this is
identical to that in the $b$-quark axis definition, namely
Eqs.~(\ref{AFBfull},~\ref{ES}).  It is worth noting that the difference
between the two definitions is the same size and in the same direction
as at $\cO(\as)$, leading to an overall difference of 0.8\% for
$\as \sim 0.12$.

We finally recall that we include an arbitrary factor $W_E$ in front of
the four-$b$ contribution to account for the way in which it is treated
in the experimental analyses.  For a fully inclusive definition, in
which each $b$ quark contributes once, $W_E$ should be set equal to 2,
while for an exclusive definition, $W_E$ should be set equal to 1.  Our
final result for the non-singlet component of the forward--backward
asymmetry, is then:
\begin{equation}
  \label{AFBfullwe} \hspace*{-1cm} A_{FB, NS}^{(2)}(W_E) \equiv
  \frac{\sigma_{A, NS}(W_E)}{\sigma_{S, NS}(W_E)} =
  A_{FB}^{(2);\mbox{\footnotesize finite}} - A_{FB}^{(0)} \left[
  (1-\smfrac12W_E) \left( 2\int E_A - \int E_S \right) + \smfrac12W_E
  \int E_S \right] \;, \hspace*{-1cm}
\end{equation}
where $A_{FB}^{(2);\mbox{\footnotesize finite}}$ is given in
Eqs.~(\ref{AFBbCNTdef},~\ref{AFBTCNTdef}) and Tables~\ref{table1}
and~\ref{table2}, $\int E_S$ is given in Eq.~(\ref{ES}), and (see
Appendix~B of Ref.\cite{CSAfb})
\begin{eqnarray}
\label{EASintb}
\hspace*{-1cm}
2\int E_A - \int E_S &=& \left(\frac{\as}{2\pi}\right)^2
C_F(C_F-\smfrac12C_A)\Bigl(0.3620\pm0.0007\Bigr),\quad\mbox{quark
axis}, \\
\label{EASintT}
\hspace*{-1cm}
2\int E_A - \int E_S &=& \left(\frac{\as}{2\pi}\right)^2
C_F(C_F-\smfrac12C_A)\Bigl(0.1144\pm0.0009\Bigr),\quad\mbox{thrust
axis}.
\end{eqnarray}
Note that the combinations of $E$-term contributions in
Eqs.~(\ref{EASintb}) and (\ref{EASintT}) are finite in the massless
limit (see the discussion in Appendix~B of Ref.\cite{CSAfb}).

Putting all these numbers together, and setting $N_f=5$, we write the
forward--backward asymmetry according to the two definitions as:
\begin{eqnarray}
  \hspace*{-1cm}
  A_{FB, NS}^{(2);b}(W_E) &=& A_{FB}^{(0)} \left[
  1-0.318\as-0.973\as^2+W_E\as^2\left(0.00405\ln\frac{Q^2}{m_b^2}-0.0240\right)
  \right],
  \hspace*{-1cm}
  \nonumber\\
  \\
  \hspace*{-1cm}
  A_{FB, NS}^{(2);T}(W_E) &=& A_{FB}^{(0)} \left[
  1-0.284\as-0.676\as^2+W_E\as^2\left(0.00405\ln\frac{Q^2}{m_b^2}-0.0233\right)
  \right],
  \hspace*{-1cm}
  \nonumber\\
\end{eqnarray}
with $\as \equiv \as(Q^2)$.  Note that the logarithmically-enhanced
term, $\ln Q^2/m_b^2$, is present for any physical ($W_E>0$) value of
$W_E$.

Putting in an explicit value for $\as$, we summarize the total QCD
correction according to the various available calculations in
Table~\ref{tab:concl}.
\begin{table}
 \begin{center}
  \hspace*{-1cm}
  \begin{tabular}{|c|c|c|c|c|}
    \hline
    & AL \cite{AL} & RvN \cite{RvN} & Our Calculation & Our Calculation \\
    & quark axis & quark axis & quark axis & thrust axis \\
    \hline
    Correction, $A_{FB}^{(2)}/A_{FB}^{(0)}$ &
    0.962 &
    0.952 &
    0.952 &
    0.956 \\
    \hline
  \end{tabular}
  \hspace*{-1cm}
 \end{center}
 \caption{\it Total QCD correction to the forward--backward asymmetry in the
    small-mass limit, with $\as=0.12$.  In each case, the thrust axis
    definition is used for the $\cO(\as)$ correction and the definition
    shown is used for the $\cO(\as^2)$ correction, as discussed in the
    text.}
 \label{tab:concl}
\end{table}
We continue to neglect all terms that vanish in the massless limit.
Since in the existing experimental analyses (see for example
Ref.\cite{abbaneo}), the known $\cO(\as)$ correction for the thrust
axis definition was included, together with the Altarelli and Lampe
quark axis value for the $\cO(\as^2)$ corrections, we do the same in
Table~\ref{tab:concl}.

We find that the difference between the Ravindran and van Neerven
calculation and ours is numerically irrelevant, being smaller than
$10^{-4}$ for $b$ quarks and $\sim2.5\!\times\!10^{-4}$ for $c$ quarks.
Therefore at the numerical precision required by current or any foreseen
experiments, we agree with their result~-- the difference is only one of
principle.  The difference between the Altarelli and Lampe calculation
and ours for the quark axis definition is more significant though, at
around 1\%.  However, the error in their calculation and the effect of
using the thrust axis definition partially cancel, and the total
difference is around 0.6\%.

We should also mention the important fact, discussed in
Ref.\cite{abbaneo}, that the experimental procedures introduce a bias
towards more two-jet-like events.  This actually decreases the size of
the QCD corrections considerably, so our number should be considered as
an upper bound on the final difference.

In Ref.\cite{CSAfb} we try to estimate the remaining uncertainties in
the forward--backward asymmetry, bearing in mind that while the 2\%
precision of current experiments is close to their final limit, a future
linear collider could be capable of experimental errors on the
left--right forward--backward asymmetry of order 0.1\%\cite{NLC}.  We
found several sources of uncertainty that all contribute at the few per
mille level.  While this is certainly sufficient for the current
precision of the data, matching the precision of a future linear
collider measurement could be extremely difficult.  It is likely that
this could only be done by making even more stringent two-jet cuts in
order to work in a region in which the corrections and their
uncertainties are smaller.

\newpage
\section*{Acknowledgements}
We are grateful to Guido Altarelli, Klaus M\"onig and especially Willy
van Neerven for discussions of the forward--backward asymmetry.

\end{document}